\documentstyle[twocolumn,epsfig,prb,aps]{revtex}

\begin{document}

\twocolumn[\hsize\textwidth\columnwidth\hsize\csname
@twocolumnfalse\endcsname

\title{Phase separation in effective hard-core boson and triplet
models in one and two dimensions}

\author{Jos\'e A. Riera}
\address{
Instituto de F\'{\i}sica Rosario, Consejo Nacional de
Investigaciones
Cient\'{\i}ficas y T\'ecnicas, y Departamento de F\'{\i}sica,\\
Universidad Nacional de Rosario, Avenida Pellegrini 250,
2000-Rosario, Argentina}
\date{\today}
\maketitle
\begin{abstract}
Effective models of hard-core hole pair bosons and triplets are
derived from the $t$-$J$ model in ladders and on the square lattice by
performing a change from site to dimer basis. The Hilbert space is
truncated by projecting out single-occupied electron states and only
nearest neighbor interactions are retained. The
resulting effective models in one and two dimensions are studied by
numerical techniques. In both spatial dimensions, the main result is
that each hole pair is surrounded by a singlet cloud expelling triplet 
excitations from its vicinity. It is suggested an interpretation
of this feature as a phase separated state between a hole-pair rich
singlet phase and an undoped triplet phase with antiferromagnetic 
correlations. The possible relevance of this result to other 
theoretical scenarios and experimental results is discussed.
\end{abstract}


\vskip2pc]

\section{Introduction}

The SO(5) theory\cite{zhang} is a very appealing framework
in which antiferromagnetic (AF) and 
superconducting (SC) 
phases of high-T$_c$ superconductors are naturally related.
This theory not only gives an explanation of the resonant peak 
observed in the SC phase\cite{respeakorig} but also predicts that
it smoothly connects with the magnon peak of the AF phase as it 
was recently observed experimentally.\cite{bourges} Another of 
the predictions of this theory, i.e. the presence of AF order
inside the vortices present when a magnetic field is applied on
the SC phase, has recently motivated a flurry of experimental 
activity\cite{lake,hoffman} confirming partially this prediction. 
At the realm of the SO(5) theory lies the fact of singlet pairing
of electrons. This is a deeply strongly correlated electrons
feature which shares with resonant valence bond (RVB)
theories\cite{rvb} and excludes an explanation of the most
important features of SC cuprates, not only the AF and SC
phases but also the intervening pseudogap phase, via Fermi
liquid or Fermi liquid instability concepts. Although there
have been many attempts of formulating a SO(5) symmetric 
model on lattices,\cite{so5ladder,so5ladd2} a connection
between the continuum theory and a more microscopic model
like the $t-J$ model is still missing.

In this sense, one of the motivations of the present study is
to help bridge the gap between the more phenomenological
SO(5) theory and the microscopic $t-J$ model. More specifically,
our goal is to obtain and study an effective model obtained
by a change of basis from the site, spin-1/2 electrons, basis
to the dimer basis, and then projecting out single occupied
dimer states. This dimer basis also indicates a relation with
RVB scenarios, in particular with its nearest neighbor 
version.\cite{kivrokset} However, in the present work the 
change of basis is performed on a single spatial dimer covering
and hence the problem of an overcomplete basis which
affects RVB models is avoided. In addition, the elimination
of single-occupied dimers implies that the excitations of
our model will not be spinons and holons, resulting from the
fractionalization of electrons, but, as in the SO(5) theory,
collective states of singlet binding of electrons in 
dimers.\cite{zhang} As a matter of fact, a recent study of the 
spin-1/2 Heisenberg model in two dimensions (2D), starting
from a singlet RVB ``soup",
has precisely arrived at the conclusion
that the elementary excitations of the AF ground state are
spin-1 excited dimers instead of spinons.\cite{eder}

The pseudogap phase\cite{tallon}, to many a central piece of the
puzzle of understanding high-T$_c$ superconductivity, is a
candidate to be a realization of a SO(5) symmetric state or
an RVB state. An effective model on dimers could then be a good
starting point to understand many features of the pseudogap phase.
In addition, the stripe phase\cite{stripes}, which in some 
cuprates appears inside the pseudogap phase, could be explained
in terms of a particular ordering of dimers that is pinned by a
spatially anisotropic exchange or by coupling to the
lattice.\cite{sachdev,sushkov} In fact, the concept of electron
singlets play a central role in one of the theories of the
stripe phase.\cite{emerykiv}

There is a more specific, recent experimental result which could
be possibly described in terms of an effective model on dimers.
A scanning tunneling microscopy study of 
Bi$_2$Sr$_2$CaCu$_2$O$_{8+\delta}$ has shown a 
phase separation between superconducting islands inside a 
percolating background with physical properties resembling those
of the pseudogap phase.\cite{granular}

Other motivations to study effective models are both technical
and conceptual. From the methodological point of view the reduction
of the Hilbert space is an advantage to most numerical techniques,
exact diagonalization, density-matrix renormalization-group, and
quantum Monte Carlo. Conceptually, the effective model may, as
it is shown below, make evident some properties which are somewhat
hidden in the original $t$-$J$ Hamiltonian.

As said above, our approach to SO(5) concepts is to map the 2D
$t$-$J$ model onto an effective model in a basis of singlet 
electron dimers. Since the original model excludes double-occupied
sites, this connection is actually closer to a recently
proposed variant of the SO(5) model called the ``projected" SO(5)
(or pSO(5)) model.\cite{pso5,dorneich} This pSO(5) model is an
attempt to close the bridge with a microscopic model on a lattice
from the other end with respect to what is intended in the present
work. Although the basis set of the pSO(5) model is identical to
that of the models that will be shown below, there are important
differences in the corresponding Hamiltonians.

The paper is organized as follows. In Section~\ref{ladsect}, an
effective model is derived for the $t$-$J$ model on ladders. In
this case, the procedure can be carried on in a much cleaner way than 
for the $t$-$J$ model on the square lattice. In addition, the 
one-dimensional (1D) character of this effective model allows its
study by exact diagonalization techniques. The main results are
common to the corresponding ones for the most interesting case, i.e.
on the square lattice, which are obtained
in Section~\ref{2dsect}. Finally, in the Conclusions, the possible
relevance of the present results to understand a number of previous 
theoretical problems as well as experimental results is discussed.

\section{Ladders}
\label{ladsect}

Although the main interest in connection with high-T$_c$ 
superconductivity is to study the interplay between various
strongly correlated electron phases in 2D, it is instructive to
start with the analysis of two-leg ladders.\cite{drs} It should be 
noticed that the $t-J$ model on this lattice has been initially
studied, among various reasons, as a realization of an RVB
state.\cite{rice} In addition, extensive studies have revealed strong 
similarities between the behaviors obtained on ladders and on
the 2D square lattice.

The Hamiltonian of the $t-J$ model is:
\begin{eqnarray}
{\cal H} = - \sum_{ \langle { i j} \rangle,\sigma } t_{ i j}
({\tilde c}^{\dagger}_{ i\sigma}
{\tilde c}_{ j\sigma} + h.c. )
+ \sum_{ \langle { i j} \rangle } J_{ i j}
({\bf S}_i \cdot {\bf S}_j -
{\frac{1}{4}} n_{i} n_{j} )
\label{ham_tJ}
\end{eqnarray}
\noindent
where the notation is standard. On ladders, $t_{ i j}=t$, $J_{ i j}=J$
along the legs, and $t_{ i j}=t_\perp$, $J_{ i j}=J_\perp$ on the rungs.
On chains and on the square lattice, the
isotropic and homogeneous case ($t_{ i j}=t$, $J_{ i j}=J$) will be
considered.

\begin{figure}
\begin{center}
\epsfig{file=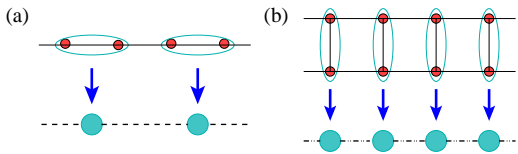,width=6.8cm,angle=0}
\end{center}
\caption{Change from a site to a dimer basis in (a) a chain and (b)
a ladder.}
\label{chbasis1}
\end{figure}

At half-filling, the exact change of basis from the site to the
dimer basis (Fig.~\ref{chbasis1}(b)) leads to the following Hamiltonian:
\begin{eqnarray}
{\cal H}_{J,ladder}^{(dimer)} = &~&2 J \sum_{\langle i,j\rangle}
(t_{0,i}^\dagger t_{0,j}^\dagger - t_{+,i}^\dagger t_{-,j}^\dagger -
 t_{-,i}^\dagger t_{+,j}^\dagger + h.c.)
\nonumber   \\
&+& 2 J \sum_{\langle i,j\rangle,\alpha} (t_{\alpha,i}^\dagger
t_{\alpha,j}+ h.c.)
+ 2 J \sum_{\langle i,j\rangle}{\bf S}_i \cdot {\bf S}_j 
\nonumber   \\
&+& J_\perp \sum_{i} (n_{t,i} -1)
\label{exch_lad}
\end{eqnarray}
\noindent
where  $t_{\alpha,i}^\dagger$ is a creation operator of a triplet
with $S^z=0, 1, -1$ ($\alpha=0,+,-$) at dimer $i$, 
$n_{t,i}=n_{0,i}+n_{+,i}+n_{-,i}$,
$n_{\alpha,i}=t_{\alpha,i}^\dagger t_{\alpha,i}$. The first term
corresponds to the spontaneous creation of two triplets out of
the vacuum with the constraint of keeping total $S^z=0$ and the
second term is a triplet ``hopping". It has been noted that this
constraint is not satisfied in the pSO(5) Hamiltonian.\cite{toy}
The third term is the usual
spin-1 Heisenberg exchange Hamiltonian. This Hamiltonian (or
equivalent expressions) has been derived previously (e.g.
Ref.~\onlinecite{so5ladd2}, and references therein) and used in 
restricted Hilbert space
diagonalizations.\cite{rungbasis} The Hilbert space of the effective
model turns out to be disconnected into two pieces: the set with
even and the set of odd number of triplets. For example, in the
subspace of total $S^z=0$ the first set is generated by successive
applications of the Hamiltonian to the initial state 
$(\ldots sssss \ldots)$, and the second one to the initial state
$(\ldots sst_0ss \ldots)$, where $s$ stands for a singlet dimer.
This splitting of the Hilbert space does not appear in the
effective model obtained for chains (Fig.~\ref{chbasis1}(a)).

\begin{figure}
\begin{center}
\epsfig{file=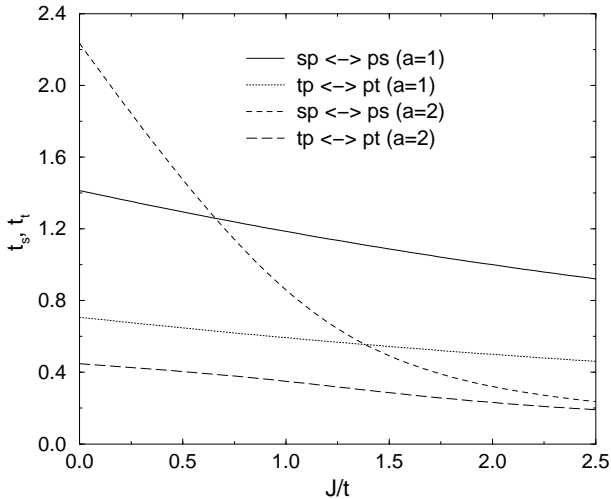,width=6.8cm,angle=-90}
\end{center}
\caption{Hopping amplitudes between pairs $p$ and singlets $s$
($t_s$) and between hole pairs and triplets  $t$ ($t_t$), as 
defined in Eq.~(\ref{hop_lad}), as a function of $J$ and for 
$a=1$ and 2. Basis change on a ladder (Fig.~\ref{chbasis1}(b)).}
\label{coef_lad}
\end{figure}

Away from half-filling, the states with a single hole occupancy on a
dimer are projected out. If ${\cal P}$ is the projection operator on 
the subspace of retained states and ${\cal Q }$ is the projection
operator on the subspace of the eliminated states, then the effective
Hamiltonian is given by the standard formula:
\begin{eqnarray}
{\cal H}_{eff} = {\cal P H P} - {\cal P H Q}\frac{1}{{\cal Q H Q}-E_0}
{\cal Q H P}
\label{project}
\end{eqnarray}
where ${\cal H} \Psi_0 = E_0 \Psi_0$, and 
${\cal H}_{eff} {\cal P} \Psi_0 = E_0 {\cal P} \Psi_0$. If ${\cal H}$
is the one given by Eq.~(\ref{ham_tJ}) on the $2\times 2$ ladder
with 2 holes, the
effective Hamiltonian will contain nearest neighbor (NN) terms only. 
In this way, the hopping term of the effective Hamiltonian, 
${\cal H}_{t,eff}$ results:
\begin{eqnarray}
{\cal H}_{t,ladder}^{(eff)} = &-& t_s \sum_{\langle i,j\rangle}
(b_j^\dagger b_i + h.c.  +n_{p,j} n_{s,i} + n_{s,j} n_{p,i})
\nonumber   \\
&-& t_t \sum_{\langle i,j\rangle,s} (b_j^\dagger t_{j,s}
t_{i,s}^\dagger b_i + h.c. +n_{p,j} n_{t,i} + n_{t,j} n_{p,i})
\nonumber   \\
&+& J_\perp \sum_{i} (n_{p,i} -1)
\label{hop_lad}
\end{eqnarray}
\noindent
where  $b_{i}^\dagger$ is a creation operator of a hole pair in the
dimer $i$, $n_{p,i}=b_{i}^\dagger b_{i}$. $n_{s,i}=1,0$ if the
site is empty (i.e., a singlet) or occupied (by a triplet or a pair)
respectively. The first term corresponds to the hopping of a pair to
a singlet site, while the second term corresponds to the hopping to
a site occupied by a triplet.
The coupling constants $t_s, t_t$ are complicated 
functions of the original parameters $\{ J, J_\perp, t, t_\perp \}$
but always $t_s > t_t $, as shown in Fig.~\ref{coef_lad} for two 
values of the lattice anisotropy $a=t_\perp /t$ ($t_s = 2t_t $,
for the isotropic ladder $a=1$). This is true even in the presence
of a NN pair-pair Coulomb repulsion. This relation
between the hopping parameters already suggests one of the most
important results of this work. That is, the pairs would tend to
be surrounded by singlets rather than by triplets in order to
gain kinetic energy. In addition, the Heisenberg term in 
Eq.~(\ref{exch_lad}) would favor the clustering of triplets.
These two combined effects would imply the phase separation 
between a pair-doped singlet phase and an undoped antiferromagnetics
phase.

Finally, the full effective Hamiltonian in the projected dimer basis
is given by:
\begin{eqnarray}
{\cal H}_{ladder}^{(eff)} = {\cal H}_{J,ladder}^{(dimer)} +
{\cal H}_{t,ladder}^{(eff)}
\label{hamhst}
\end{eqnarray}

The 1D effective model can be studied by exact diagonalization 
techniques (Lanczos algorithm). Most of the results below were 
obtained on a $L=12$ chain with periodic boundary conditions.
Some computations for a $L=16$ chain show that 
finite size effects are not important. All results shown
below correspond to fixed number of pairs
$N_p$, and were obtained for the ground state, ${\bf k}=(0,0)$.
In the following, all energies and coupling constants are
expressed in units of $t$ of the original $t$-$J$ Hamiltonian
Eq.~(\ref{ham_tJ}). 

\begin{figure}
\begin{center}
\epsfig{file=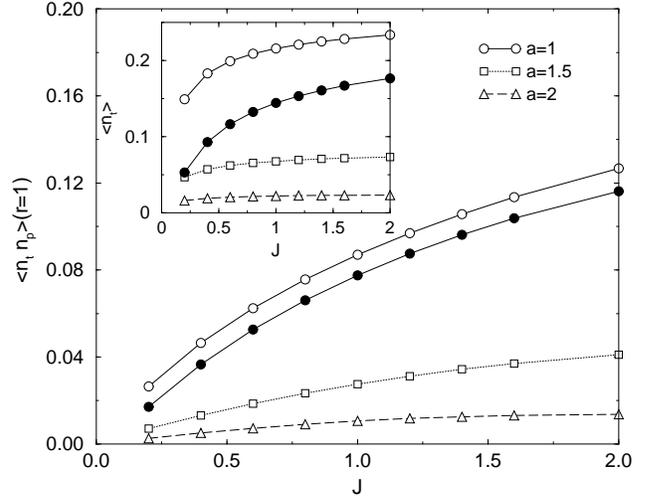,width=6.8cm,angle=-90}
\end{center}
\caption{Triplet-pair correlation at $r=1$, $L=12$, $N_p=1$ (open
symbols), as a 
function of $J$ and for the values of the anisotropy ratio $a$
indicated on the plot. The inset shows the triplet density for the
same cluster and parameters. Results for $N_p=2$ (filled circles),
$a=1$ are also included.}
\label{triplad1}
\end{figure}

In Fig.~\ref{triplad1}, the correlation between a pair and 
a triplet, $\langle n_t n_p \rangle$ at distance $r=1$ (NN 
sites) is shown for the $L=12$ chain with one pair, as a 
function of $J/t$ and for three ladder anisotropy ratios. 
It can be seen that the probability of finding a triplet 
next to a pair is much smaller than the probability of finding
a pair next to a singlet. The later, in the one pair system, is 
simply $\langle n_s n_p \rangle=1-\langle n_t n_p \rangle$.
It is more important the result that, as it can be easily
seen in Fig.~\ref{triplad1}, $\langle n_t n_p \rangle~(r=1)$ 
is always smaller than the triplet density $\langle n_t \rangle$ 
(inset) for the same $J/t$. The normalization of the later is 
such that $n_s+n_t+n_p=1$, $n_p=0.0833$ (one pair), and
$n_p=0.1667$ (two pairs). This result suggests that triplets 
are expelled from the vicinity of a pair.

\begin{figure}
\begin{center}
\epsfig{file=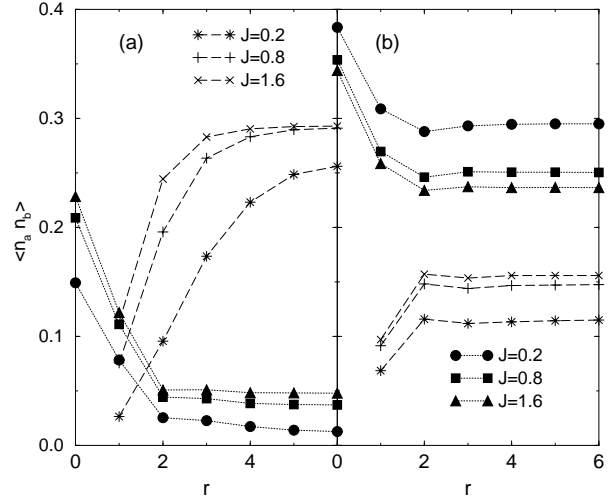,width=6.8cm,angle=-90}
\end{center}
\caption{Diagonal correlations as a function of distance
obtained for $L=12$, $N_p=1$, $a=1$, and various values of $J$, as
indicated on the plot.
(a) Triplet-triplet (dot lines) and triplet-pair (dashed lines)
correlations; (b) singlet-singlet (full symbols, divided by 2) and 
singlet-triplet (dashed lines) correlations.}
\label{corrlad1}
\end{figure}

The complete picture can be inferred from Fig.~\ref{corrlad1} by
looking at various 
correlations $\langle n_a n_b \rangle$, ($a,b=t_0, t_+, t_-,p$)
as a function of distance. From now on, the study will be limited to
the isotropic case ($t=t_\perp$, $J=J_\perp$ in the original
$t$-$J$ model) but similar results were also found for all
$a > 1$ investigated. In this Figure, which corresponds also
to the system with one pair, it can be
observed that the triplet-triplet correlation is maximum at 
$r=0$, while the triplet-pair correlation is maximum at the 
largest distance on the chain. That is, triplets try to stay
as far apart as possible form a pair.
On the other hand, although singlet-singlet correlations are
also maximum at $r=0$, the singlet-pair correlations are 
maximum at $r=0$. The final piece is that the singlet-triplet
correlation is maximum at the maximum distance. Similar 
results were obtained for the larger $L=16$ site chain.

The picture emerging from these correlations is that pairs move
preferentially in a background of singlets, and that both
pairs and singlets try to keep themselves away from triplets.
Thus, the system is separated between a pair-rich singlet phase
(it is tempting to consider this phase as a doped ``RVB" phase, 
even though that the present effective models are obtained for
a single dimer covering) and a pair-poor triplet-rich phase
which can be identified
as an undoped AF phase.

\begin{figure}
\begin{center}
\epsfig{file=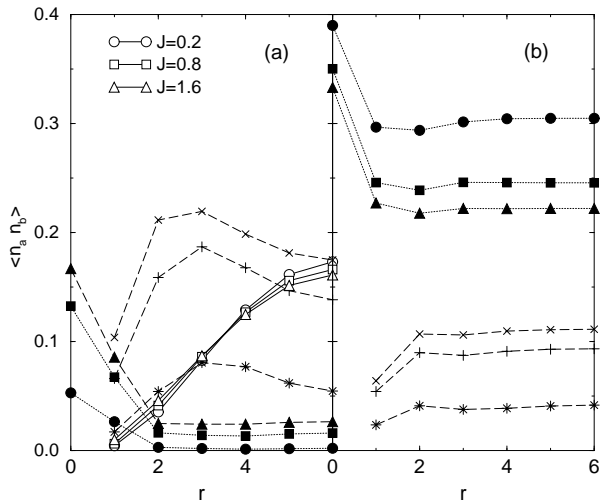,width=6.8cm,angle=-90}
\end{center}
\caption{Diagonal correlations as a function of distance
obtained for $L=12$, $N_p=2$, $a=1$, and various values of $J$ as
indicated on the plot.
(a) Triplet-triplet (full symbols), pair-pair (open symbols) and 
triplet-pair (dashed lines)
correlations; (b) singlet-singlet (full symbols, divided by 2) and 
singlet-triplet (dashed lines) correlations.}
\label{corrlad2}
\end{figure}

Essentially the same behavior is observed in the case of two
pairs present in the system. Results for the triplet-pair
correlation in NN sites and the triplet density as a function
of $J/t$ are also included in Fig.~\ref{triplad1} for comparison
with the one pair case. Both the probability of 
finding a pair next to a triplet and the triplet density are
smaller than for the one pair case. A likely explanation of
this behavior is that the introduction of more pairs in the
system increases the volume of the pair-doped ``RVB" phase
leaving less room for the AF phase.

In Fig.~\ref{corrlad2}, diagonal correlations are shown for 
$L=12$ and same parameters as in Fig.~\ref{corrlad1}. The new
feature in Fig.~\ref{corrlad2} with respect to Fig.~\ref{corrlad1}
is the fact that two pairs repel themselves as can be read
from the fact that pair-pair correlations are maximum at the
largest distance. Then, the same qualitative behavior found in
the one-pair case holds: the system separates into a pair-doped
singlet region, here formed by two islands and a pair-poor 
triplet-rich region, presumably with short-range AF order, in
this case filling the space between those two islands. Notice
also in Fig.~\ref{corrlad2}(b) the singlet-singlet correlations
falling down to its bulk value within a lattice spacing.

\section{Two dimensions}
\label{2dsect}

\begin{figure}
\begin{center}
\epsfig{file=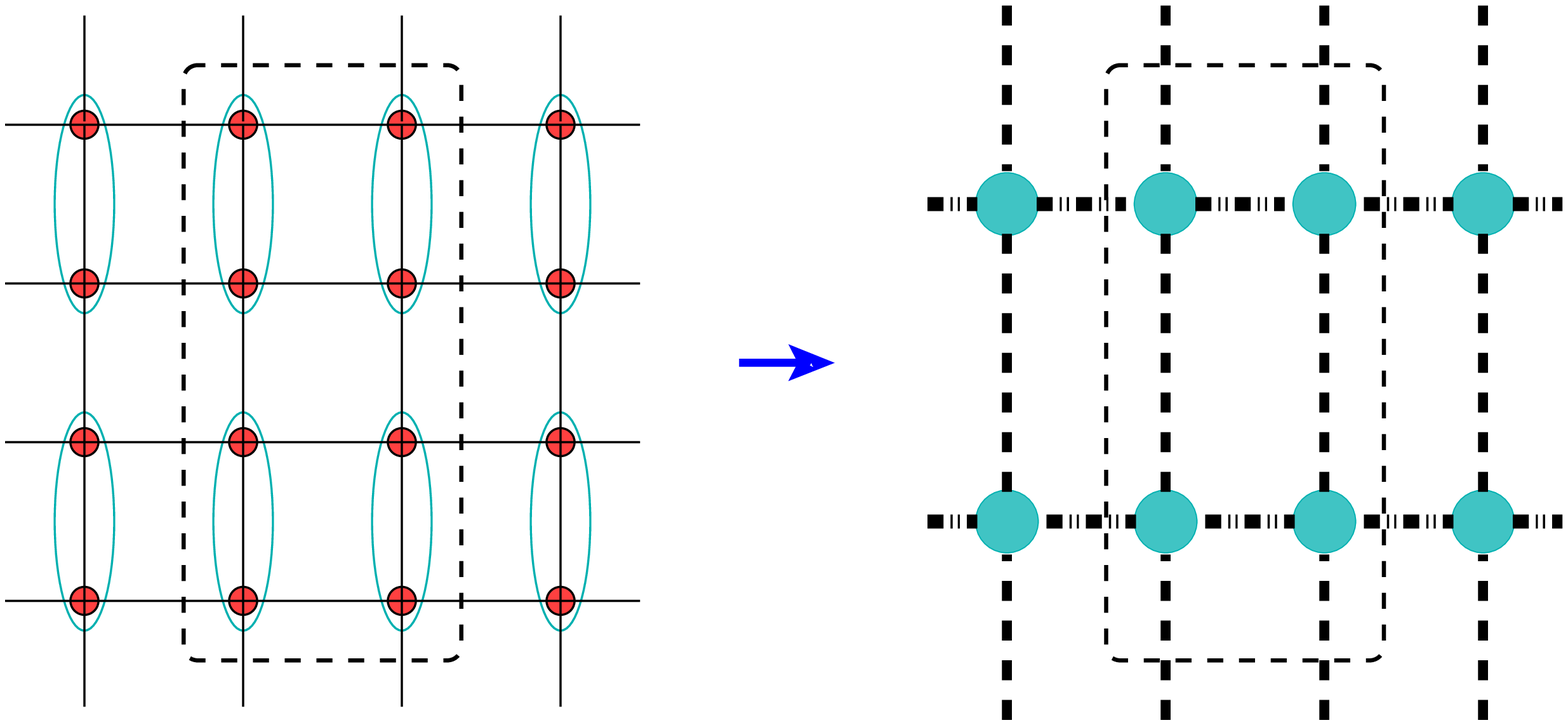,width=6.8cm,angle=0}
\end{center}
\caption{A possible change from a site to a dimer basis in the
square lattice. Dotted and dash-dotted lines indicate effective
magnetic interactions in the new basis.}
\label{chbasis2}
\end{figure}

The most important situation is that of the square lattice, which
corresponds to the CuO$_2$ planes in superconducting cuprates.
There are again infinitely many different ways in which a change
from the site to the dimer basis can be performed.
One of them is shown in Fig.~\ref{chbasis2}.
A well-known feature of choosing a dimer basis like the one
depicted in Fig.~\ref{chbasis2} is that the rotation
invariance of the square lattice is broken and it is not simple
to restore it at the level of the effective Hamiltonian. In the
ladder case, examined in the previous Section, this is not 
important since the lattice itself is spatially anisotropic.
In the case of the square lattice, the purpose of the present
study is to provide indications of the presence, in a
rotationally-broken effective model for the square lattice, of 
the singlet-AF phase separation
already observed in ladders
and to suggest that this feature should be also present
in a rotational invariant formulation.\cite{altman}

At half-filling the change of basis implies an exact mapping of the
original spin-1/2 Heisenberg model onto a model written in terms
of triplets and singlets. In the basis change indicated in 
Fig.~\ref{chbasis2} the resulting model is anisotropic with
ladder-like interactions (Fig.~\ref{chbasis1}(b)) given by 
Eq.~(\ref{exch_lad}) in the horizontal direction and chain-like 
interactions
(Fig.~\ref{chbasis1}(a)) in the vertical direction, which are
given by the following Hamiltonian:
\begin{eqnarray}
{\cal H}_{J,chain}^{(dimer)} = &-& J \sum_{\langle i,j\rangle}
(t_{0,i}^\dagger t_{0,j}^\dagger - t_{+,i}^\dagger t_{-,j}^\dagger -
 t_{-,i}^\dagger t_{+,j}^\dagger + h.c.)
\nonumber   \\
&+& J \sum_{\langle i,j,\alpha \rangle} 
(t_{\beta,i}^\dagger t_{\gamma,j}^\dagger t_{\alpha,i} -
t_{\gamma,i}^\dagger t_{\beta,j}^\dagger t_{\alpha,i} + h.c.)
\nonumber   \\
&-& J \sum_{\langle i,j\rangle,\alpha} (t_{\alpha,i}^\dagger
t_{\alpha,j}+ h.c.)
+ J \sum_{\langle i,j\rangle}{\bf S}_i \cdot {\bf S}_j 
\nonumber   \\
&+& J_\perp \sum_{i} (n_{t,i} -1)
\label{exch_lin}
\end{eqnarray}
\noindent
where $\beta= +, 0, -$, $\gamma = -, +, 0$ for $\alpha= 0, +,-$
respectively.
The sign of the second term depends on the definition of the singlet.
It should be noticed that this term, as in the ladder case, locally
conserves the 
total $S^z$. This Hamiltonian can be read, with a slightly different
notation though, in Ref.~\onlinecite{eder}, and alternative or
similar derivations can be found in several other
places.\cite{sachdev,rungbasis}
Finally, the exchange part of the Hamiltonian in the dimer basis is:
\begin{eqnarray}
{\cal H}_{J,square}^{(dimer)} = {\cal H}_{J,ladder}^{(dimer)} +
{\cal H}_{J,chain}^{(dimer)}
\label{exch_2d}
\end{eqnarray}
\noindent
with the ladder (chain) term acting on the horizontal (vertical)
direction as indicated in Fig.~\ref{chbasis2}.

\begin{figure}
\begin{center}
\epsfig{file=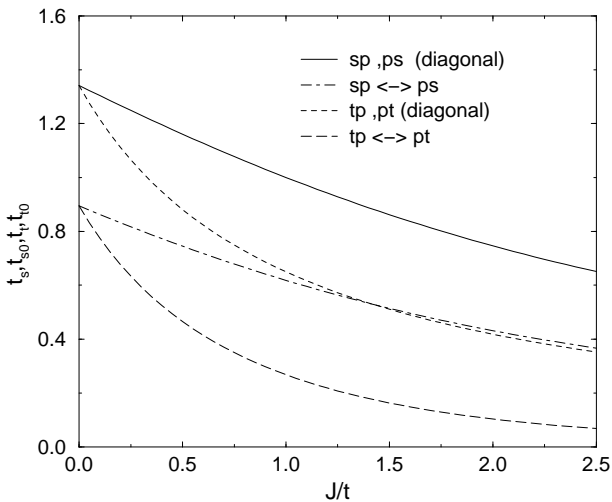,width=6.8cm,angle=-90}
\end{center}
\caption{Hopping amplitudes between pairs $p$ and singlets $s$
($t_s$,$t_{s0}$) and between pairs and triplets  $t$
($t_t$,$t_{t0}$), as defined in Eq.~(\ref{hop_lin}), as a function
of $J$. Basis change on a chain (Fig.~\ref{chbasis1}(a)).}
\label{coef_lin}
\end{figure}

Let us consider the original $t$-$J$ Hamiltonian defined on a
chain. To obtain the hopping part of the effective model,
the projection defined by Eq.~(\ref{project}) is again used.
In this case, ${\cal{H}}$ is the $t$-$J$ Hamiltonian on a
4-site chain (Fig.~\ref{chbasis1}(a)) with two holes and the 
effective one
will contain NN interactions only. The resulting effective
hopping term is very similar to the one in Eq.~(\ref{hop_lad}):
\begin{eqnarray}
{\cal H}_{t,chain}^{(eff)} = 
- t_s^\prime \sum_{\langle i,j\rangle} (b_j^\dagger b_i + h.c.)
\nonumber   \\
- t_{s0} \sum_{\langle i,j\rangle} (n_{p,j} n_{s,i} + n_{s,j} n_{p,i}) 
\nonumber   \\
- t_t^\prime \sum_{\langle i,j\rangle,s} (b_j^\dagger t_{j,s}
t_{i,s}^\dagger b_i + h.c.)
\nonumber   \\
-t_{t0} \sum_{\langle i,j\rangle} (n_{p,j} n_{t,i} + n_{t,j} n_{p,i})
\nonumber   \\
+ J_\perp \sum_{i} (n_{p,i} -1)
\label{hop_lin}
\end{eqnarray}
\noindent
As in the case of Eq.~(\ref{hop_lad}), the hopping amplitudes
satisfy $t_s^\prime > t_t^\prime$, and $t_{s0} > t_{t0}$ (Fig.
\ref{coef_lin}) again
favoring the movement of pairs away from triplet-rich regions.

Coming back to the square lattice, the correct procedure is to
to take ${\cal{H}}$ as the $t$-$J$ Hamiltonian on the eight-site
cluster indicated by a dashed box in Fig.~\ref{chbasis2}. As
a result, the effective Hamiltonian contains three- and four-site
terms in addition to NN interactions. In order to keep the
Hamiltonian as close as possible with the proposed pSO(5)
model,\cite{pso5,dorneich} only NN hopping interactions as
given by Eqs.~(\ref{hop_lad}) and (\ref{hop_lin}) in the horizontal
and vertical directions respectively are retained. The signs of
these hopping terms coming from the eight-site cluster calculation
are the same as in Eqs.~(\ref{hop_lad}) and (\ref{hop_lin}). To
compensate for neglecting three- and four-site terms, the hopping
amplitudes in (\ref{hop_lad}) are re-scaled by a single constant
$\alpha$ and the amplitudes in (\ref{hop_lin}) by another
constant $\beta$. A reasonable fit of the energies of the effective
model on the $4 \times 2$ cluster to the exact energies of the
$t$-$J$ model on the $4\times 4$ cluster in the whole range
studied, $0 \le J \le 2.5$, is achieved with $\alpha=1$ and
$\beta=0.5$.
In this range of $J$, the relative difference between these two
energies is less than $0.01$.
The exchange part of the effective model is given by
Eq.(~\ref{exch_2d}).

A first insight on the properties of the effective model can be
gained by studying the $4 \times 4$ cluster. Due to the large 
dimension of the Hilbert space
($\approx 7.76 ~10^7$ for $N_p=1$ and $\approx 1.50 ~10^8$ for
$N_p=2$ taking into account translational invariance), conventional
exact diagonalization techniques cannot be applied, except by
resorting to massive computers. Alternatively, a diagonalization
in a systematically expanded Hilbert space (SEHS)\cite{sehs} is
used. With a number of states $\approx 5~10^6$, variational energies
within $1\%$ of the exact energies, estimated by extrapolating to the 
full dimension of the Hilbert space, are obtained.
In order to study larger clusters, a quantum Monte Carlo(QMC) 
technique with the conventional worldline checkerboard
decomposition\cite{reger} is used. Although there are no fermions
involved, there is a ``minus sign problem" which makes impossible
the study at
low temperatures. There are several terms in the effective 
Hamiltonian which lead to ``minus sign" configurations in the
2+1-dimensional space. An important reduction of this problem
is achieved by not generating those configurations 
with interacting cubes which do not conserve the parity of 
the number of triplets on its top and bottom plaquettes. Of
course, the QMC algorithm is no longer exact but nevertheless
it provides a reasonable approximation  to its exact
behavior. By eliminating certain transitions, it might be possible
that certain regions of the phase space are disconnected. To
cope with this problem, at each temperature, results were averaged 
over  at least four independent runs starting from different initial
states. Variations in the values of the energy from different runs
were somewhat larger than the statistical error of each
run but nevertheless smaller than $1\%$ in all cases.
In addition, since the 
the present study concerns zero temperature properties, the
simulations are restricted to the subspace of zero
total magnetization and only local moves are included in the
algorithm.
Overall
simulation error bars are approximately twice the size of the
symbols used.

\begin{figure}
\begin{center}
\epsfig{file=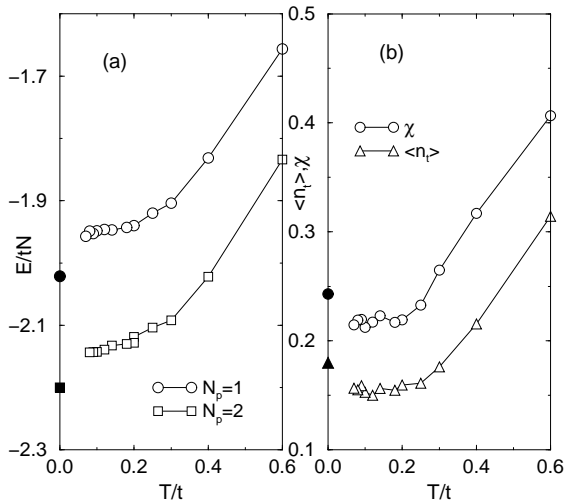,width=6.8cm,angle=-90}
\end{center}
\caption{(a) Variation of energy per site with temperature in the 
$S^z=0$ subspace in the $4\times 4$ cluster with one and two pairs,
$J/t=0.8$. (b) Variation of triplet density and $(\pi,\pi)$
magnetic structure factor with temperature in the subspace $S^z=0$
in the $4\times 4$ cluster with two pairs,$J/t=0.8$. In both panels,
solid symbols at $T=0$ indicate the corresponding values obtained by
SEHS diagonalization.}
\label{truncMC}
\end{figure}

Evolution of various quantities with temperature in the
$S^z=0$ subspace on the $4\times 4$ cluster, $J=0.8 t$, with one
and two pairs are shown in Fig.~\ref{truncMC}. The corresponding 
zero temperature results obtained by diagonalization in an 
expanded Hilbert space
are also included for comparison. As expected the energies
obtained by the approximated QMC technique are higher than the
obtained by diagonalization (which are virtually exact), while 
the triplet densities are smaller than the exact ones.

\begin{figure}
\begin{center}
\epsfig{file=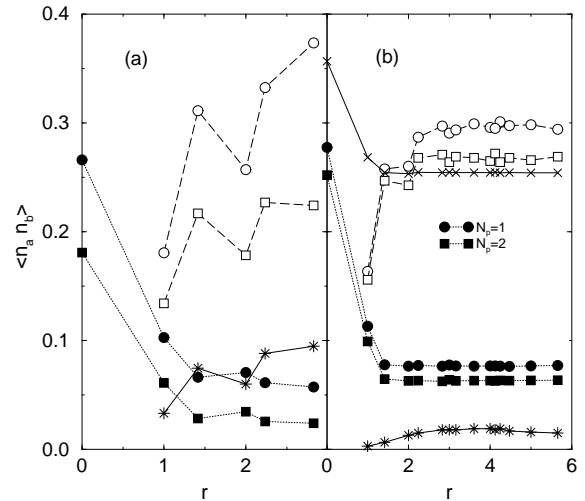,width=6.8cm,angle=-90}
\end{center}
\caption{Triplet-triplet (full symbols) and triplet-pair (open
symbols) correlations as a function of distance,
$N_p=1, 2$, $J=0.8 t$, (a) on the $4\times 4$ cluster (SEHS) and
(b) on the $8\times 8$ one (QMC, $\rm T=0.09 t$).
Pair-pair (stars) and singlet-singlet (crosses, (b) only)
correlations for $N_p=2$ are also shown.}
\label{cortrunmc}
\end{figure}

The same correlations previously studied on ladders are shown
in Fig.~\ref{cortrunmc}(a) for the $4\times 4$ cluster, with one
and two pairs, $J=0.8 t$. These ground state correlations,
obtained by SEHS at ${\bf k}=(0,0)$, have been averaged
over the two-directions $x$ and $y$. The most important result is
again that the triplets try to locate as far as possible from the
pairs. Similar results were obtained for the other correlations
shown in Fig.~\ref{corrlad2}, and for all values of $J/t$ examined.

Results on the $8\times 8$ cluster obtained by QMC technique
at $\rm T=0.09 t$ for one and
two pairs are shown in Fig.~\ref{cortrunmc}(b). The behavior of
triplet-triplet and triplet-pair correlations and pair-pair
corrrelations are qualitatively similar to those found in the
smaller cluster at zero temperature. These results are very
suggestive that triplets are expelled from the vicinity of a
pair. Notice also that, as for ladders (Fig.~\ref{corrlad2}),
singlet-singlet correlations decay very rapidly to their bulk 
value.

\begin{figure}
\begin{center}
\epsfig{file=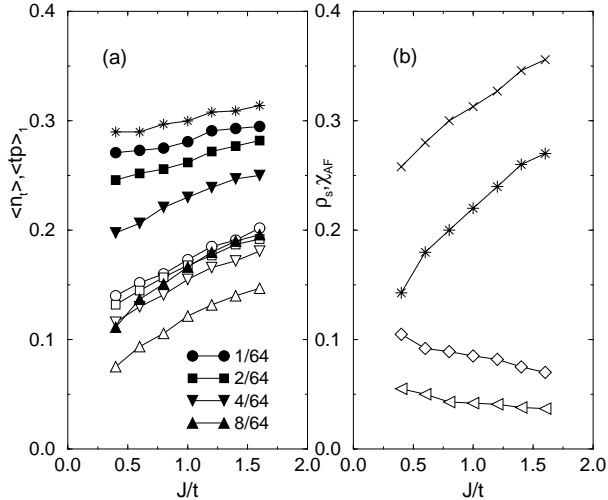,width=6.8cm,angle=-90}
\end{center}
\caption{Results obtained by QMC on the $8\times 8$ cluster, 
$\rm T=0.09 t$ at various fillings and as a function of $J/t$.
(a) Triplet density (full symbols) and triplet-pair correlations
at NN sites (open symbols). The triplet-pair correlation at the
maximum distance for $N_p=1$ are shown with stars(b) AF
structure factor (triangles: $N_p=4$, diamonds: $N_p=8$) and 
superfluid density (plus: $N_p=4$, stars: $N_p=8$).}
\label{totall}
\end{figure}

A more systematic study as a function of $J/t$ leads to
the results depicted in Fig.~\ref{totall}(a). In all cases,
the probability of finding a triplet near a pair (open symbols)
is smaller than the average probability of finding a triplet on a
given site (full symbols). Consistently with the idea of pairs
expelling triplet excitations, the probability of finding a 
triplet at the largest distance of a pair (only shown for the 
one pair case) is larger than the average triplet probability.
Singlet-singlet correlations vs. distance

Since the system contains charged pair bosons, it is important to
calculate the superfluid density as a measure of superconducting
properties in this model. To calculate the superfluid density it 
The procedure to calculate the superfluid density\cite{batrouni}
starts by computing the correlation:
\begin{eqnarray}
C(\tau)=\langle \Delta p_x(\tau) \Delta p_x(0) \rangle
\label{cortau}
\end{eqnarray}
\noindent
where
$\Delta p_x(\tau)=p_x(\tau +1)-p_x(\tau)$ and 
$p_x(\tau)=\sum x(i,\tau)$, where $x(i,\tau)$ is the $x$-coordinate
of pair $i$ at imaginary time $\tau$ and the sum extends over all
the pairs in the system. The superfluid density follows from:
\begin{eqnarray}
\rho_s \approx \lim_{\omega \rightarrow 0} \int_0^\beta 
d\tau C(\tau) exp(-i \omega \tau)
\label{fourier}
\end{eqnarray}
\noindent
It is well-known that this quantity vanishes if the simulation is
carried in a zero-winding number subspace which is the case in the
present study. The way out of this problem
stems from the fact that the winding number may be nonzero in
{\em half} of the Trotter, imaginary time, direction, being also
nonzero but with an opposite sign in the other half. The 
superfluid density then would come out by taking the Fourier 
transform in (\ref{fourier}) between zero and $\beta /2$. This
procedure should be exact in the limit of infinite Trotter number
which is the limit in which on the other hand the whole worldline
algorithm is valid.
This procedure has been thoroughly checked in the hard-core boson
model where exact and numerical results are
available.\cite{hebert,bernardet} 

The superfluid density, $\rho_s$, and the magnetic structure factor
at $(\pi,\pi)$, $\chi_{AF}$, on the $8\times 8$ cluster with four
and eight pairs, are shown in Fig.~\ref{totall}(b) as a function of
$J/t$.
Although the ultimate dominance of SC or AF, or eventually the
coexistence of both SC and AF orders can be determined only by a
finite size extrapolation which is out of the scope of the
present study, two qualitative features are apparent. First, as a
pair doping is increased there is an enhancement of SC and a 
suppression of AF. Second, as $J/t$ is increased $\chi_{AF}$ also
increases while $\rho$ decreases. Similar results were obtained
with an even simplified model which mimics the spin triplets by spin
doublets\cite{toy} and can be obtained by projecting out the $S^z=0$
component of the triplets, $t_0$, again by using (\ref{project}).
The mutually exclusive behavior of AF and superconductivity, which
has been shown both in the present work and previously in
Ref.~\onlinecite{toy}, can now be understood in terms of this real
space phase separation between SC-singlet and AF regions. The growth
of one of each phases reduces the space available for the other
phase.

\section{Conclusions}

In the first place, the present work is concerned with
the mapping of the $t$-$J$ model into an effective
model of pairs and triplets moving on a ``sea" of singlets. The
strategy adopted was first to map exactly the magnetic interactions
in the undoped system and then to determine the effective hopping
interactions in the two-hole sector of the original Hamiltonian.
This second procedure is an approximate one because it implies 
projecting out single-occupied dimers and truncating 
the range of the interactions nearest neighbors. In the case of
ladders this
procedure is reasonably clean. The choice of dimers corresponds 
to the strong coupling limit which extends virtually to the 
anisotropic case. An effective model in one dimension is obtained
in which the hopping couplings between pairs and singlets are
larger than between pairs and triplets.

In the square lattice the procedure to obtain the effective model
is more complicated. In the first place, the simple dimer covering
adopted breaks rotational invariance of the lattice. Still, 
the mapping of the magnetic interactions at half-filling
is exact. In the 
second place, even by restricting the range of hoppings to 
nearest neighbors, three- and four-site interactions appear at
the effective level. To eliminate these interactions another
fit to the energies of a small $t$-$J$ cluster was performed
to rescale the effective hopping constants.
The resulting effective hopping couplings show the same behavior
as those obtained for the ladder case. 

The resulting effective models were studied by numerical techniques,
exact diagonalization in the case of the model obtained for the
ladder lattice and approximate diagonalization and Quantum Monte
Carlo for the square lattice case. Again the results obtained 
in the effective model for ladders are more clear than the ones
obtained for the square lattice.

From the study of several correlations the following picture 
emerges. Pairs are surrounded predominantly by singlets, and 
triplet excitations are located as far apart as possible. This is 
a kind of phase separation between a pair-rich ``RVB" phase
and a pair-poor
triplet-rich phase which would correspond to a phase with at least 
short-range interactions (since triplet excitations restore AF
order from a RVB state\cite{eder}) and would quite likely be 
insulating. Pair-pair correlation functions indicate that
pairs try to be situated at the maximum distance (this behavior
is more definite in ladders than in 2D). This behavior might 
correspond to pairs moving in a singlet phase macroscopically
separated from the triplet phase. Alternatively, the phase
separation might be microscopic:  pair-singlet islands moving
in a triplet background. The behavior of triplet-pair
correlations in 1D could favor this second scenario. In any case,
these PS scenarios provide an explanation for the AF-SC mutual 
exclusion shown in the previous Section.

Further studies are necessary to distinguish between both types 
of phase separation. In the macroscopic PS scenario it is expected
a superfluid density comparable with that of the hard-core boson
model while on the microscopic PS scenario it would be considerable
reduced. On theoretical grounds, the emerging PS picture could be 
realted to the AF-SC coexistence phase predicted by 
SO(5) theories. The exclusion of triplets and pairs is highly
nontrivial taking into account that previous studies on the
$t$-$J$ model\cite{rieradag} have suggested a bound state between
a d$_{x^2-y^2}$ pair and a triplet. However, the internal structure
of a pair, essential in the analysis of Ref.~\onlinecite{rieradag},
is lost in the present study.\cite{siller}
Finally, it
is also tempting to relate this PS state to recent observations of
inhomogeneities in Bi$_2$Sr$_2$CaCu$_2$O$_{8+\delta}$
(Ref.~\onlinecite{granular}. It might be
possible that out-of-plane negative Coulomb centers could attract
and pin pair-singlet islands in their surroundings.

\begin{acknowledgments}
The author wishes to acknowledge many useful discussions with
Prof. S. Maekawa and to Prof. M. Boninsegni for suggesting the
procedure to compute the superfluid density.
\end{acknowledgments}

\end{document}